\documentclass[%
 aip,
 amsmath,amssymb,
 reprint,%
]{revtex4-1}

\usepackage{graphicx}
\usepackage{dcolumn}
\usepackage{bm}

\usepackage[utf8]{inputenc}
\usepackage[T1]{fontenc}
\usepackage{mathptmx}

\usepackage{amssymb}
\usepackage{graphicx}
\usepackage{caption}
\usepackage{subcaption}
\usepackage{multirow}

\usepackage{kantlipsum}
\usepackage{amsmath}
\usepackage{amsfonts}
\usepackage{amssymb}
\usepackage{dcolumn} 
\usepackage{braket}
\usepackage{booktabs}
\usepackage{multirow}
 \usepackage{movie15}
\usepackage{hyperref}

\usepackage{times}
\usepackage{epsfig}
\usepackage{xcolor}
\usepackage{tikz}
\usepackage{float}
\usepackage{color}
\usepackage{inputenc}
\usepackage{booktabs}
\usepackage{pgfplotstable}
\usepackage{subcaption}

\def\bx {{\bf x}}

\def\bkappa {\boldsymbol{\kappa}}

\newcommand{\beq}{\begin{equation}}
\newcommand{\eeq}{\end{equation}}
\newcommand{\ba}{\begin{eqnarray}}
\newcommand{\ea}{\end{eqnarray}}

\font\myfont=cmr12 at 15pt

\begin{document}

\preprint{AIP/123-QED}

 \title[Topological guided waves within non-hexagonal structures]{Experimental observations of topologically guided water waves within non-hexagonal structures }


\author{Mehul P. Makwana}
 \email{mm107@ic.ac.uk}
 \affiliation{Department of Mathematics, Imperial College London, London SW7 2AZ, UK}
\affiliation{Multiwave Technologies AG, 3 Chemin du Pre Fleuri, 1228 Geneva, Switzerland}
 
 \author{Nicolas Laforge}%
 \email{nicolas.laforge@femto-st.fr}
 \affiliation{Institut FEMTO-ST, CNRS UMR 6174, Université de Bourgogne Franche-Comté, 25000 Besançon, France}

 \author{Richard V. Craster}%
 \affiliation{Department of Mathematics, Imperial College London, London SW7 2AZ, UK}
 \affiliation{UMI 2004 Abraham de Moivre-CNRS, Imperial College, London SW7 2AZ, UK}
 
 \author{Guillaume Dupont}%
\affiliation{Aix Marseille Univ, CNRS, Centrale Marseille, IRPHE UMR 7342, 13013 Marseille, France}

 \author{S\'ebastien Guenneau}%
 \affiliation{UMI 2004 Abraham de Moivre-CNRS, Imperial College, London SW7 2AZ, UK}
 
 \author{Vincent Laude}%
\affiliation{Institut FEMTO-ST, CNRS UMR 6174, Université de Bourgogne Franche-Comté, 25000 Besançon, France}

 \author{Muamer Kadic}%
\affiliation{Institut FEMTO-ST, CNRS UMR 6174, Université de Bourgogne Franche-Comté, 25000 Besançon, France}

\date{\today}

\begin{abstract}
We investigate symmetry-protected topological water waves 
  within a strategically engineered square lattice system. Thus far, symmetry-protected topological modes in hexagonal systems have primarily been studied in electromagnetism and acoustics, i.e. dispersionless media.
    Herein, we show experimentally how crucial geometrical properties of square structures allow for topological transport that is ordinarily forbidden within conventional hexagonal structures. We perform numerical simulations that take into account the inherent dispersion within water waves and devise a topological insulator that supports symmetry-protected transport along the domain walls. Our measurements, viewed with a high-speed camera under stroboscopic illumination, unambiguously demonstrate the valley-locked transport of water waves within a non-hexagonal structure. Due to the tunability of the energy's directionality by geometry our results could be used for developing highly-efficient energy harvesters, filters and beam-splitters within dispersive media.  

\end{abstract}

\maketitle

\begin{figure}
\includegraphics[width=7cm]{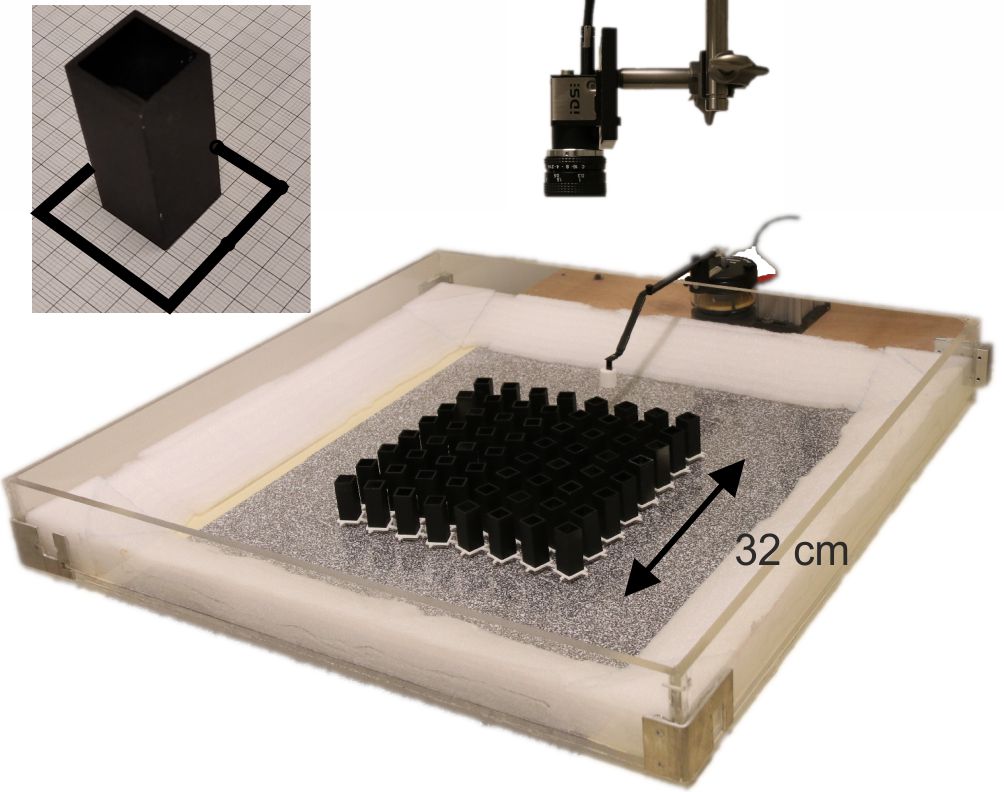}
\caption{Experimental setup: A crystal is assembled using square shaped aluminum tubes 7~cm in height arranged in a square array with different orientations using a plastic positioning frame at the bottom of the tank ($80 \times 80$ cm$^2$ with 60-degree oblique edges made of soft polystyrene to mimic PMLs). A mechanical straight paddle holding a small plastic cylinder is used to generate water waves. The tank is continuously illuminated and images of water waves are recorded with a high speed camera placed on the top. A black and white random pattern is placed under the tank to provide water elevation measurement using an image cross-correlation algorithm. The experimental setup was inspired by the work of Moisy et al; \cite{Moisy2009}.}
\label{setup}
\end{figure}

Considerable recent activity in wave phenomena is motivated through topological effects and focused on identifying situations where topological protection occurs that can enhance, or create, robust wave guidance along edges or interfaces. Remarkably, the core concepts that gave rise to topological insulators,  originating within quantum mechanics \cite{kane_z2_2005} carry across, in part, to classical wave systems\cite{khanikaev_two-dimensional_2017, ozawa_topological_2019}. 
Topological insulators can be divided into two broad categories: those that preserve time-reversal symmetry (TRS), and those which break it. We concentrate upon the former due to the simplicity of their construction that solely requires passive elements. By leveraging the discrete valley degrees of freedom, arising from degenerate extrema in Fourier space, we are able to create robust symmetry-protected waveguides. These valley states are connected to the quantum valley-Hall effect and hence this research area has 
been named valleytronics\cite{behnia_polarized_2012, schaibley_valleytronics_2016}. 

Hexagonal structures are the prime candidates for valleytronic devices as they exhibit symmetry induced Dirac cones at the high-symmetry points of the Brillouin zone (BZ); when perturbed  these Dirac points can be gapped, leading to well-defined $KK'$ valleys distinguished, from each other, by their opposite chirality or pseudospin. 

This pseudospin has been used in a wide variety of dispersionless wave settings to  design valleytronic devices \cite{dong_valley_2017, xiao_valley-contrasting_2007}. Here we extend the earlier research by examining a highly-dispersive physical system, i.e. water waves 
 and move away from hexagonal structures. The topological protection afforded by these valley states is attributed to, both, the orthogonality of the pseudospins as well as the Fourier separation between the two valleys \cite{makwana_designing_2018-1}. The vast majority of the valleytronics literature, inspired by graphene, opts to use hexagonal structures \cite{he_acoustic_2016, schomerus_helical_2010, ye_observation_2017, cheng_robust_2016, wu_direct_2017, xia_topological_2017, qiao_electronic_2011, makwana_geometrically_2018, tang_observations_2019, proctor_manipulating_2019}. However a negative that emerges with these, especially when dealing with complex topological domains \cite{makwana_designing_2018-1}, is that certain propagation directions are restricted due to mismatches in chirality between incoming and outgoing modes. Notably, this has led to hexagonal structures being prohibited from partitioning energy in more than two-directions \cite{qiao_electronic_2011, cheng_robust_2016, wu_direct_2017}. 

In this Letter, we  demonstrate experimentally how a strategically designed square structure also allows for the emergence of valley-Hall edge states 
as well as allowing for the excitation of modes that are not ordinarily ignited within hexagonal valley-Hall structures. Additionally, the system chosen differs from the vast majority of the earlier literature \cite{he_acoustic_2016, schomerus_helical_2010, ye_observation_2017, cheng_robust_2016, wu_direct_2017, xia_topological_2017, qiao_electronic_2011, makwana_geometrically_2018, tang_observations_2019, proctor_manipulating_2019} that has focused on an idealised situation in which the dispersion of the host medium has been avoided. This assumption restricts the applicability of the earlier studies to a small subset of, potentially useful, physical platforms that could host topological effects.
Most notably, this assumption does not hold for water wave systems, which generally support highly dispersive surface waves\cite{lighthill_waves_1978}.  The combination of topological physics applied to water waves is a relatively unexplored area \cite{wu_topological_2018, laforge_observation_2019}; those that have conducted experiments have either focused on $1$D systems\cite{wu_topological_2018} or the hexagonal valley-Hall structure \cite{laforge_observation_2019}. Potential applications of this budding area include controlling ocean wave energy\cite{bennetts18a}, in a non-intrusive manner, for energy-harvesting or erosion mitigation\cite{Dupont17}.

The fluid within our domain has a constant depth of $h=4$~cm and contains a periodic array of rigid, vertical and bottom mounted, square objects (2~cm of side length in a 4~cm square array) that perforate the free surface of the liquid (see the experimental set-up in Fig.~\ref{setup}). The planar coordinates are denoted by ${\bx} = (x_1, x_2)$ whilst the vertical upward direction has the coordinate $x_3$ ; the origin is prescribed to be at the mean free surface. Under the usual assumptions of linear water wave theory\cite{lighthill_waves_1978}, where the fluid is assumed to be inviscid, incompressible with irrotational flow, there exists a velocity potential $\Phi$ \cite{linton90a} such that
 \beq
 \Phi(\bx, t)=\Re e \left [ \phi({\bx})\frac{\cosh(k(x_3+h))}{\cosh(k h)}\exp(-i\omega t) \right ],
\eeq
where $\omega$ denotes the angular frequency. The wavenumber, $k$ the real positive solution of the 
 dispersion relation
 \beq
\left(g k +\frac{\sigma}{\rho}k^3\right)\tanh(k h)=\omega^2,
\label{eq:water_dispersion}
\eeq
 is used as a proxy for the frequency;\cite{laforge_observation_2019} in Eq. \eqref{eq:water_dispersion} $g=9.81$~m~s$^{-2}$ is gravitational acceleration, $\sigma=0.07$~N~m$^{-1}$ is the surface tension between air and water and $\rho=10^3$~kg~m$^3$ the water density. Then $\phi$, the reduced potential,  satisfies the Helmholtz equation,
 \beq
 \left( \nabla_{\bx}^2 + k^2 \right)\phi({\bx}) = 0,
\eeq
where this equation holds at the mean free surface and the subscript $\bx$ indicates differentiation with respect to $\bx$.
 and no-flow boundary conditions on the vertical rigid cylinders: 
 taking ${\bf n}=(n_1,n_2)$ as the unit outward normal to the square tubes' surface, $ \partial\phi/\partial {\bf n} = 0$ on each of them.

\begin{figure}[h!]
\centering
\includegraphics[width=8cm]{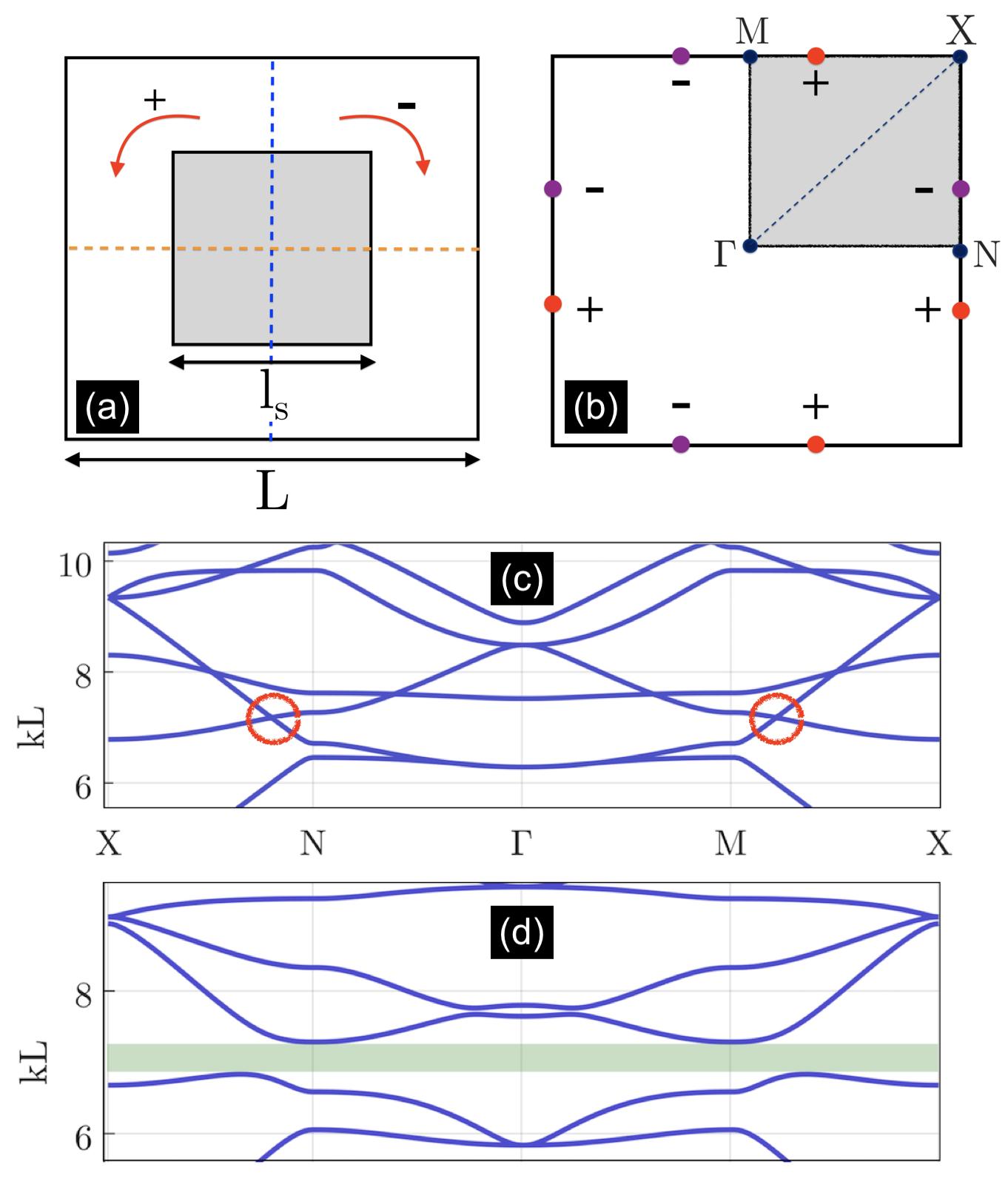}
\caption{Geometry, band structure and topological features: (a) Periodic cell (physical space) of the square lattice with sidelength $L$ showing a square inclusion of sidelength $l_s$ inside it. Mirror-symmetry breaking rotation (arrows) and lines (dashed) are also shown. (b) In reciprocal space, the points $\Gamma NX$ denote the extrema of the IBZ that we extend to $\Gamma N X M$ to show the two topologically inequivalent regions; the two distinct $\text{sgn}(C_v)$ values are indicated by $\pm$ signs around the perimeter of the BZ and these are associated with the $+$ perturbation in panel (a) (the $-$ perturbation would result in opposite $\text{sgn}(C_v)$'s, see \cite{xia_observation_2018}). The $\text{sgn}(C_v)$ positions resemble those in \cite{makwana_tunable_2019, makwana_topological_2019}. (c) Band diagram for the configuration in (a), with two circles marking the position of the strategically engineered Dirac cones
(d) Band diagram when the inclusion is rotated through an angle of $20^\circ$. A band-gap highlighted in green emerges from the symmetry breaking perturbation at Dirac points.} 
\label{bulk}
\end{figure}

When the problem is posed in terms of the reduced potential, $\phi$, as the Helmholtz equation, with periodically arranged inclusions (the tubes), this directly maps across to the phononic crystal literature. Recognising the periodicity guides us to setting $\phi(x_1,x_2)=\phi_{\boldmath\kappa}(x_1,x_2)\exp(i\bkappa\cdot{\bx})$ with $\bkappa$ as the Bloch wavenumber and $\phi_{\boldmath\kappa}$ as the periodic piece of the Bloch solution\cite{laforge_observation_2019}.
 A key ingredient, that guides the experiments, is an understanding of the   dispersion relation, relating $k$ to the Bloch-wavenumber, ${\bkappa}$ spanning the BZ ${\bkappa}\in[0;\pi/L]\times[0;\pi/L]$ (see Fig \ref{bulk}(a)) 
  for an infinite perfectly periodic square lattice system; we determine this relationship numerically.
 The geometry and band structures are shown in Fig. \ref{bulk}; for a square array, lattice constant $L$, the irreducible Brillouin zone (IBZ) is an eighth of the BZ. Despite this, we opt to plot around a quadrant of the BZ as this will incorporate the two distinct Dirac cones that are essential for our valley-Hall states. The desired quadrant has the following vertices: $X=(\pi/L,\pi/L)$, $N=(\pi/L,0)$, $\Gamma=(0,0)$, $M=(0,\pi/L)$. 

\emph{Results} \textemdash The unrotated cellular structure chosen, Fig. \ref{bulk}(a), contains, both, horizontal and vertical mirror symmetries along with four-fold rotational symmetry. Hence, in its entirety the structure has $C_{4v}$ point group symmetries. Notably, it is the presence of these mirror symmetries that yield Dirac cones along the outer edges of the BZ, Fig. \ref{bulk}(b,c) \cite{he_emergence_2015, makwana_topological_2019, makwana_tunable_2019, xia_observation_2018, manzanares-martinez_dirac_2019}. Note, that rectangular structures (wallpaper group $P2mm$) which possess these mirror symmetries will also yield these non-symmetry repelled Dirac cones \cite{makwana_tunable_2019, makwana_topological_2019}. In contrast to hexagonal structures the position of these degeneracies can be tuned by varying the geometrical or material parameters of the system \cite{xia_observation_2018}. By rotating the internal square inclusion, both mirror symmetries are broken thereby yielding the band-gap shown for the dispersion curves in Fig. \ref{bulk}(d). The residual valleys, that demarcate the band-gap, are locally imbued with a nonzero valley-Chern number \cite{ochiai_photonic_2012},
\beq
C_{v} =  \frac{i}{2\pi} \int_{S} \nabla_{\bkappa} \times \phi_{\kappa}^*(\bx)\cdot  \nabla_{\bx}\phi_{\kappa}(\bx) d\bkappa = \frac{i}{2\pi} \oint_{\gamma} \phi_{\kappa}^*(\bx) \nabla_{\bx}\phi_{\kappa}(\bx) \cdot d{\bf l} 
\label{eq:valley_Chern} \eeq
where the path integrated around ($\gamma$) encircles a particular valley and the superscript $*$ denotes the complex conjugate. Despite the calculation (and name) of the $C_v$ resembling that of its TRS breaking counterpart, namely the Chern number, there is an important difference: the former is not a quantized quantity whilst the latter is. The surface associated with $\gamma$ is not on a closed manifold hence the Gauss–Bonnet theorem \cite{Berger_differential} does not hold. Despite this, the opposite pseudospin modes have a bijective relationship with $\text{sgn}(C_v) = \pm 1$ which itself can be classed as a topological integer \cite{qian_theory_2018}. From this we can apply the bulk-boundary correspondence for certain edge terminations thereby guaranteeing the existence of valley-Hall edge states. 

\begin{figure}[htb!]
\centering
\includegraphics[width=8cm]
{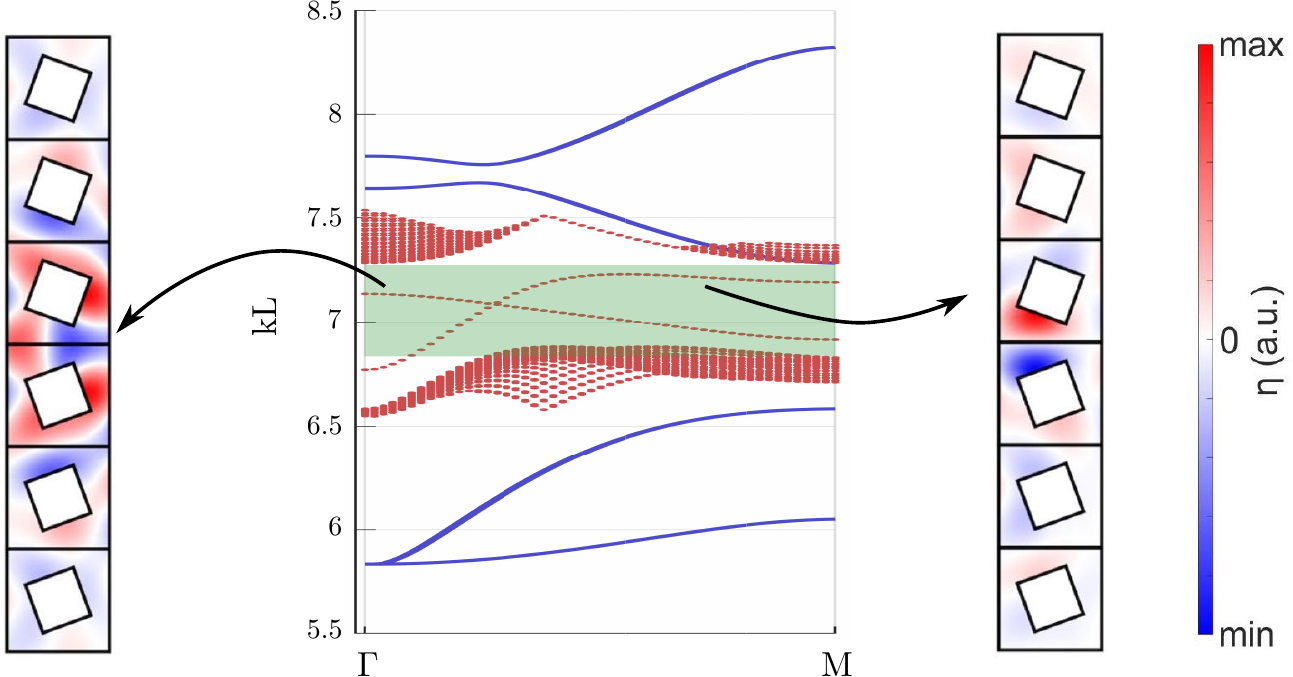}
\caption{Valley-Hall edge states: Band diagram for a ribbon with the upper/lower inclusions rotated clockwise/anti-clockwise. The real parts of the even and odd eigenmodes within the band-gap are shown (in red) as are several close-by ribbon modes (also in red); the latter's values range between their max and min and a.u stands for arbitrary units ($\eta$ is the vertical displacement of the surface). The blue curves are from Fig. \ref{bulk}(d), i.e. bulk modes along $\Gamma M$.  
Numerically, using finite elements, we take a long ribbon of $N$ inclusions, apply Dirichlet boundary conditions top and bottom and extract the modes decaying away from the interface.}
\label{ribbon}
\end{figure}

Motivated by this, we place a perturbed cellular structure, that contains a positively or negatively rotated inclusion, above its reflectional twin. This results in a pair of gapless edge modes that almost span the entirety of the band-gap, see Fig. \ref{ribbon}. Here we use ``gapless" to refer to the crossing of the concave and convex (opposite parity) modes. This distinguishes valley-Hall systems, that are topological, with those that are not and have coupled edge states; for example, the armchair termination within hexagonal structures produces gapped edge states that are, in turn, less robust \cite{fefferman_bifurcations_2016}.

The gapless nature of the states, and in turn the applicability of the Gauss–Bonnet theorem, is contingent upon the termination chosen containing projections of valleys with identical $\text{sgn}(C_v)$. Unique to this specific square structure, the different parity eigenmodes belong to the \emph{same} interface (see Fig. \ref{ribbon}), rather than different interfaces. Despite this 
both, concave and convex states, have opposite parity and hence remain orthogonal. The relationship between the interfaces arises due to the mirror-symmetry relationship between the media either side of the interface in Fig. \ref{ribbon}. This also implies that a right-propagating mode along one of the interfaces is a left-propagating mode on the other. A numerical illustration of this phenomenon is found in \cite{makwana_tunable_2019} where decaying Hermite polynomials were used to oust a specific parity edge state, along both interfaces, non-simultaneously. This phenomenon does not occur for hexagonal structures where the different parity eigenmodes belong to different interfaces. This relationship between the two interfaces allows for propagation, within our square structure, that is ordinarily forbidden within graphene-like structures. Coupling between modes, that are hosted along different interfaces, is crucial for energy navigation around sharp corners \cite{makwana_geometrically_2018} and within complex topological domains \cite{qiao_electronic_2011, cheng_robust_2016, wu_direct_2017, makwana_designing_2018-1}.
Further explanation for this phenomenon can be found in \cite{makwana_tunable_2019, makwana_topological_2019}.




\begin{figure*}[htb!]
\centering
\includegraphics[width=15cm]{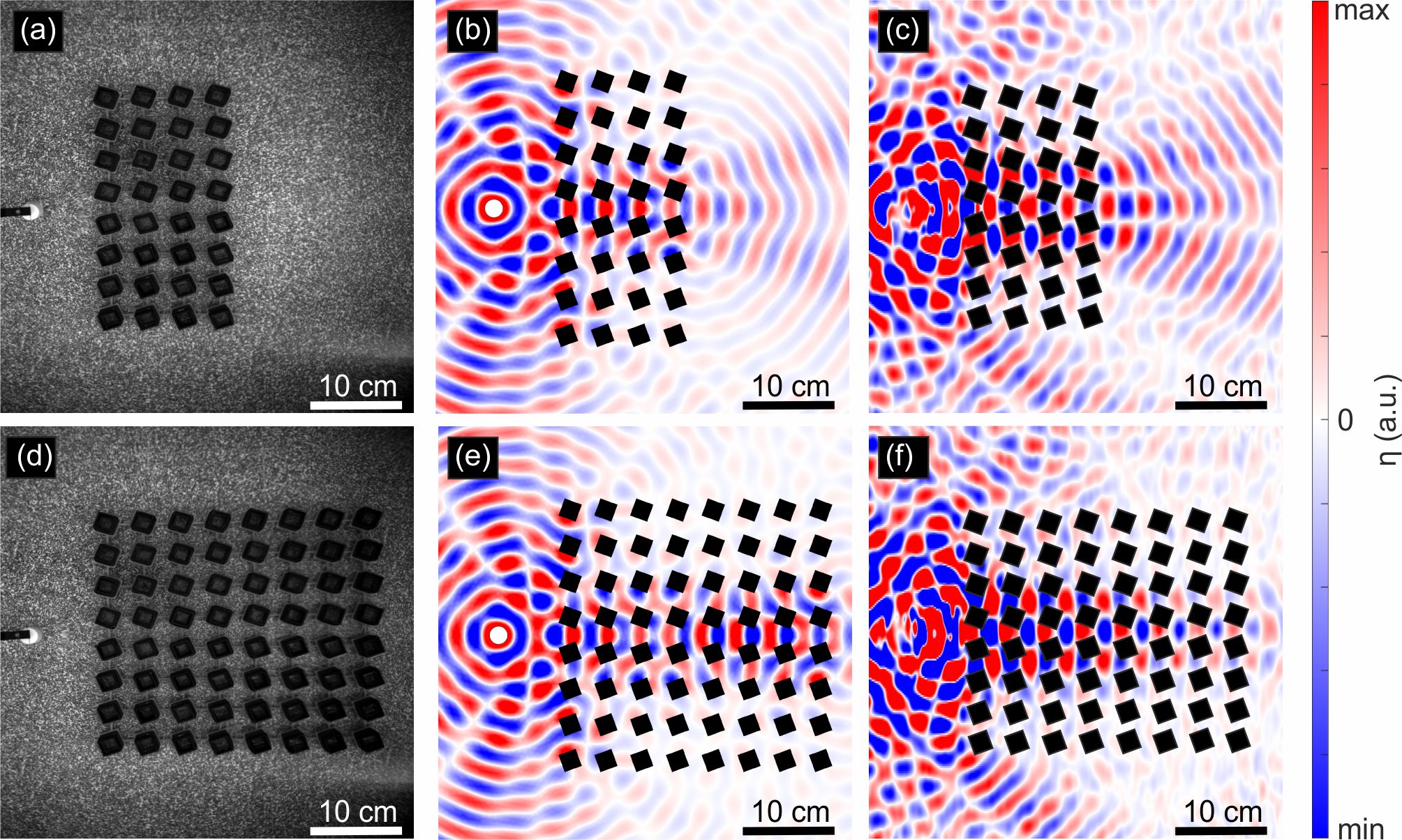}
\caption{Valley-Hall edge states: Experiments, and simulations. (a,d) Experimental set-up showing the top view of the water surface perforated by $4\times 8$ and $8\times 8$ square rigid inclusions respectively. (b,e) Corresponding numerical calculations; (c,f) Experimentally observed wavepatterns. These valley-Hall states are generated by a monopolar source operating at a frequency of 7.3~Hz and placed 6~cm from the domain wall edge and propagate between inclusions rotated by $\pm 20^\circ$.
}
\label{zlms}
\end{figure*}

The propagation of water waves is imaged at the surface of the water tank of Fig.~\ref{setup}. A mechanical paddle holding a circular cylinder is shaken at a controlable frequency. Cylindrical waves originating from the monopolar source are observed numerically and experimentally in Fig. \ref{zlms}(a, e).
The experimental setups for a topologically nontrivial interface, with two different lengths, is shown in Figs. \ref{zlms}(b, f); the upper/lower halves have square inclusions rotated clockwise/ anti-clockwise in order to break the mirror symmetries and generate the valley-edge states required. Images were acquired by a high speed camera and post processed using a cross-correlation algorithm\cite{Moisy2009}; each image was discretized into 360 areas each composed of 16 pixels.

Full-wave numerical simulations, performed using COMSOL Multiphysics (a commercial finite element scheme), for tightly confined valley-Hall edge states, Figs. \ref{zlms}(c, d), show excellent agreement with the experiments, Figs. \ref{zlms}(g, h), despite our model not taking into account contact-line effects that occur between the water and the solid pillars, viscosity or nonlinearity. These square structure valley-Hall edge states  have longer-wavelengths than their hexagonal counterparts and hence the distance between the pillars is subwavelength. A frequency modulated monopolar source is generated that ignites the even-parity valley-Hall edge state. The observed patterns are associated with the surface curvature where the coloured regions are indicative of the vertical elevation of the water level. The localisation of the topological edge state is clearly evident when comparing two interfaces of differing lengths, i.e. four or eight squares in Figs. \ref{zlms}(c, f). 
Notably, the valley-Hall state that propagates across four columns, Fig. \ref{zlms}(c), radiates almost isotropically upon exit. In the absence of any rods the energy would radiate isotropically away from the source. The broadbandedness of this effect is demonstrated via the experimental results shown in the supplementary material. The tight-confinement of these dispersive water waves, within a strategically designed square structure, is a highly nontrivial and unique observation.

\begin{figure}[H]
\centering
\includegraphics[width=8cm]{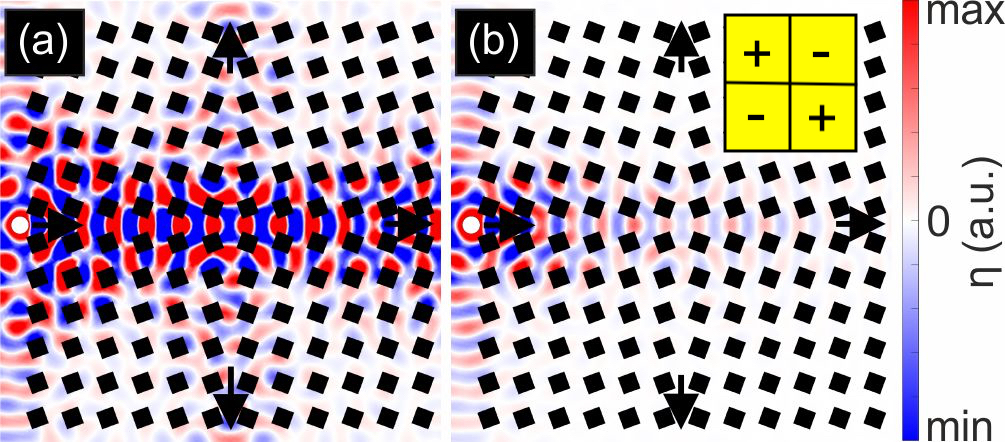}
\caption{Numerical illustration for a three-way splitter at 7.3~Hz: Four quadrants of alternately squares rotated clockwise and counter-clockwise. The valley-Hall phases, associated with the propagating modes, are shown within the yellow inset. (a) has no losses whilst (b) incorporates attenuation introduced by considering a complex wave velocity with 2\% imaginary part relative to the real part.  }
\label{three}
\end{figure}

We now strategically extend our earlier design, Figs. \ref{zlms}(b, f), to engineer four structured quadrants that results in a three-way topological energy-splitter, Fig. \ref{three}. We rotate the bottom-right and top-right inclusion sets anti-clockwise and clockwise, respectively, thereby creating four distinct domain walls upon which the valley-Hall states reside. The monopolar source triggers a wave, from the leftmost interface, into upward and downward modes along with continued rightward propagation. Incidentally, the most pronounced displacement pattern is along the two geometrically distinct horizontal interfaces. This continued rightward propagation is \emph{forbidden} for hexagonal systems \cite{qiao_electronic_2011, cheng_robust_2016, wu_direct_2017, makwana_designing_2018-1}. For coupling between the incident mode and the right-sided mode the chirality's must match and this does \emph{not} happen for hexagonal structures. Contrastingly, this mismatch in chirality is overcome for the square structure as the right-sided interface is the reflectional partner of the left-sided interface. Hence, the incident mode need only to couple to itself in order to continue it's rightward propagation. This subtle relationship between the mirror-symmetry generated Dirac cones, and the subsequent, mirror-symmetry related interfaces allows for propagative behaviour not readily found within the valleytronics literature, Fig. \ref{three}. 

In the experiments there are multiple loss mechanisms on the lengthscales we are operating at: viscous attenuation \cite{padrino07a}, contact line losses associated with the frictional drag of the meniscus moving up and down the rigid pillars i.e. contact angle hysteresis \cite{joanny84a}, Marangoni effects due to surface tension variations \cite{craster09a} and their effect on capillary-gravity waves, and then the nonlinear inertial effects that are ignored through linearisation of the Navier-Stokes equations\cite{lighthill_waves_1978}.
Inspecting Fig. \ref{three}, which is a simulation, we note that the amplitude of the   interface modes propagating up and down are about $1/10$ of that propagating left to right.
 In Fig. 5(b) we insert losses, by lumping them into a complex wave velocity, of just 2\% (we choose this to be simply illustrative and to demonstrate that losses can in an experiment obscure the subtle effects often sought in topological systems) and this reduces the signal quite dramatically in the up and down interfaces; in experiments the amplitudes were too small to be accurately measured and we attribute this to the loss mechanisms we describe above.


%
We have experimentally shown the existence of topological valley-Hall transport for gravity-capillary water waves within a non-hexagonal structure. We have also simulated a three-way topological multiplexer for the same highly-dispersive system and cautioned that  losses may lead to low amplitudes. These demonstrations open up a useful way for design in energy transport: the conventional symmetry constraints associated with hexagonal structures can be relaxed leading to richer designs of waveguides and multiplexers within highly-dispersive systems.

\emph{Supplementary Material}\textemdash
The supplementary material contains additional experimental images for valley-Hall edge states in crystals of different size. Also shown are images and simulations in the absence of the crystal. It also contains basic information about the finite element models used.

We acknowledge V. Pêcheur, E. Carry, T. Daugey and F. Chollet for the experimental part. This work has been supported by the EIPHI Graduate School (contract ANR-17-EURE-0002) and by the French Investissements d'Avenir program, project ISITE-BFC (contract ANR-15-IDEX-03).
R.V.C. and M.P.M. thank the UK EPSRC for their support through grants  EP/L024926/1 and EP/T002654/1.



%

\pagebreak
\widetext
\begin{center}
\textbf{{\myfont Supplementary Material}}
\end{center}
\setcounter{equation}{0}
\setcounter{figure}{0}
\setcounter{table}{0}
\setcounter{page}{1}
\makeatletter
\renewcommand{\theequation}{S\arabic{equation}}
\renewcommand{\thefigure}{S\arabic{figure}}
\renewcommand{\bibnumfmt}[1]{[S#1]}
\renewcommand{\citenumfont}[1]{S#1}

\section*{Simulating water waves using finite elements}
The implementation of the water wave equation within a finite element package (such as Comsol Multiphysics) is fairly straightforward. The way to proceed is as follows: We
first multiply the main equation valid at the mean free surface 
\beq
 \left( \nabla_{\bx}^2 + k^2 \right)\phi({\bx}) = 0,
\eeq
(with the subscript $\bx$ indicating differentiation with respect to $\bx$) by a smooth test function $\Psi$ with a compact support in the periodic cell $Y=[0;L]\times [0;L]$ before integrating over $Y$. Using Green’s formula, and the fact that the Neumann boundary condition on the inclusion in $Y$ is a natural condition we obtain the so-called weak form of the Helmholtz equation
\begin{equation}
 -\int_Y \nabla_{\bx}\phi({\bx})\cdot\nabla_{\bx}\Psi({\bx}) \, d{\bx}
 +k^2\int_Y \phi({\bx})\Psi({\bx}) \, d{\bx} = 0,
\end{equation}
Solving  the Floquet-Bloch  spectral  problem  amounts  to  looking  for  the  countable  set  of real  positive eigenvalues $k^2$ and associated non-zero eigensolutions $\phi_{\boldmath\kappa}({\bx})$
for a given $\boldmath\kappa$ such that
$\phi(x_1,x_2)=\phi_{\boldmath\kappa}(x_1,x_2)\exp(i\bkappa\cdot{\bx})$. Here, $\bkappa$ is the Bloch wavenumber and $\phi_{\boldmath\kappa}({\bx})$ is the periodic piece of the Floquet-Bloch solution. When we let $\bkappa$ vary within the Brillouin zone $BZ=[0;\pi/L]\times [0;\pi/L]$, we obtain the band spectrum. It is customary in condensed matter theory to only consider the edges of the BZ, and this enables us in our water wave problem to represent the band spectrum with dispersion curves, rather than dispersion surfaces, in a way similar to what was done in \cite{farhat_analytical_2008,laforge_observation_2019}.

The discrete formulation is set up with nodal elements (first-order triangular elements where the
unknowns are the values of the scalar field at the vertices of the triangles and the interpolated field is
piecewise linear in the triangles). In order to find Floquet-Bloch modes, with the finite element method, we must discretise our grid and apply discrete Bloch conditions.
A scalar discrete field on the square cell $Y$
with Floquet-Bloch conditions relates the left and the right sides, and the bottom and the top sides of $Y$, respectively. The details of this algorithm can be found e.g. in \cite{langlet_1995}. We implement the Floquet-Bloch water wave problem in comsol multiphysics using the global PDE equation form and built-in Floquet-Bloch conditions. 
We model the scattering problem using the weak form of the Helmholtz equation (S1) where the right hand side now has a forcing term (a point source). Perfectly matched layers (PMLs) provide a reflectionless interface between the region of interest (the waterwave tank of square shape when viewed from above) and the PMLs (four rectangles and four small squares surrounding the square domain) at all incident angles. PMLs were originally introduced by Berenger \cite{berenger_perfectly_1994} in electromagnetism, where the regions of PMLs consisted of anisotropic absorptive dielectric media. In our case, we consider some 'liquid PMLs' defined by the Helmholtz equation,
\begin{equation}
 \left( \nabla_{\bx}\cdot\mu\nabla_{\bx} + \rho k^2 \right)\phi({\bx}) = 0,
\end{equation}
which holds at the mean free surface of the liquid. Moreover, the matrix $\mu$ is given by
\begin{equation}
\mu=
\left(
\begin{array}{ll}
\frac{s_ys_z}{s_x}&0 \\
0 &\frac{s_xs_z}{s_y}
\end{array}
\right) \; ,
\end{equation}
and the scalar parameter 
\begin{equation}
\rho=\frac{s_xs_y}{s_z} ,
\end{equation}
where $s_x$,$s_y$ and $s_z$ represent complex stretched coordinates (as introduced by Berenger \cite{berenger_perfectly_1994}) defined as,
\begin{equation}
s_x=a+ib \; , \; s_y=s_z=1 \; ,
\end{equation}
for the absorption of water waves along the $x_1$-direction, with the roles of $x$ and $y$ interchanged for absorption of water waves along the $x_2$ direction. Moreover, for absorption of water waves along the $x_1=x_2$ direction one considers
\begin{equation}
s_x=s_y=a+ib \; , \; s_z=1 \; .
\end{equation}

\section*{Isotropic radiation in homogeneous liquid}
\begin{figure}[h!]
\centering
\includegraphics[width=8.5cm]{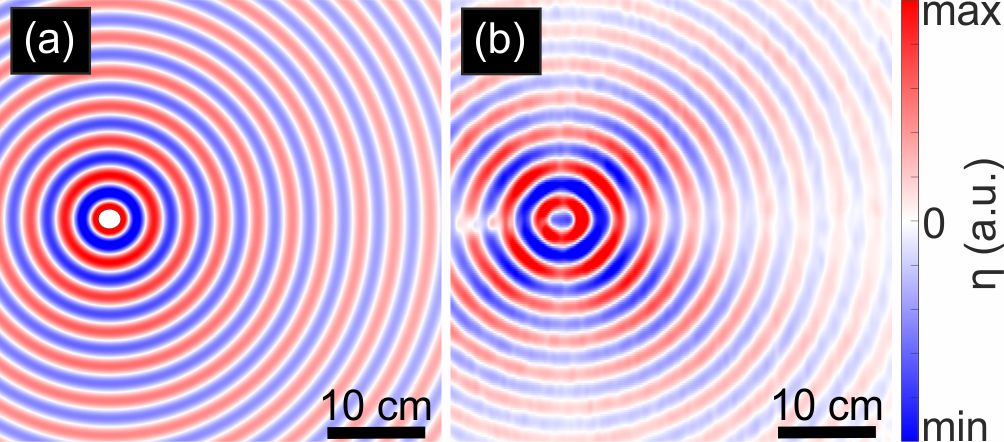}
\caption{(a) Numerical and (b) experimental propagation of cylindrical water waves from a monopolar source.}
\label{figS1}
\end{figure}

\noindent
Figure \ref{figS1}(a) shows the isotropic radiation caused by the monopolar source within a homogeneous liquid surrounded by the PMLs (not shown) to avoid any reflection on the boundary of the computational domain (taking $a=b=1$). This water pattern agrees well with an experimental measurement, see Figure \ref{figS1}(b), which demonstrates that 60-degree oblique edges made of soft polystyrene on either sides of the tank mimic the PMLs well.

We note that the measured vertical displacement of the liquid surface $\eta$ in Figure \ref{figS1} is
related to the potential $\phi$ computed numerically via
\begin{equation}
\eta(x_1,x_2,t)=\Re\{-i\frac{\omega}{g}\phi(x_1,x_2)\exp(-i\omega t)\} \; ,
\end{equation}
where $t$ is the time variable, $g$ and $\omega$ denote gravitational acceleration and angular frequency, respectively, as defined in the main article. 

\section*{Broadband domain wall states}
\begin{figure}[h]
\centering
\includegraphics[width=17cm]{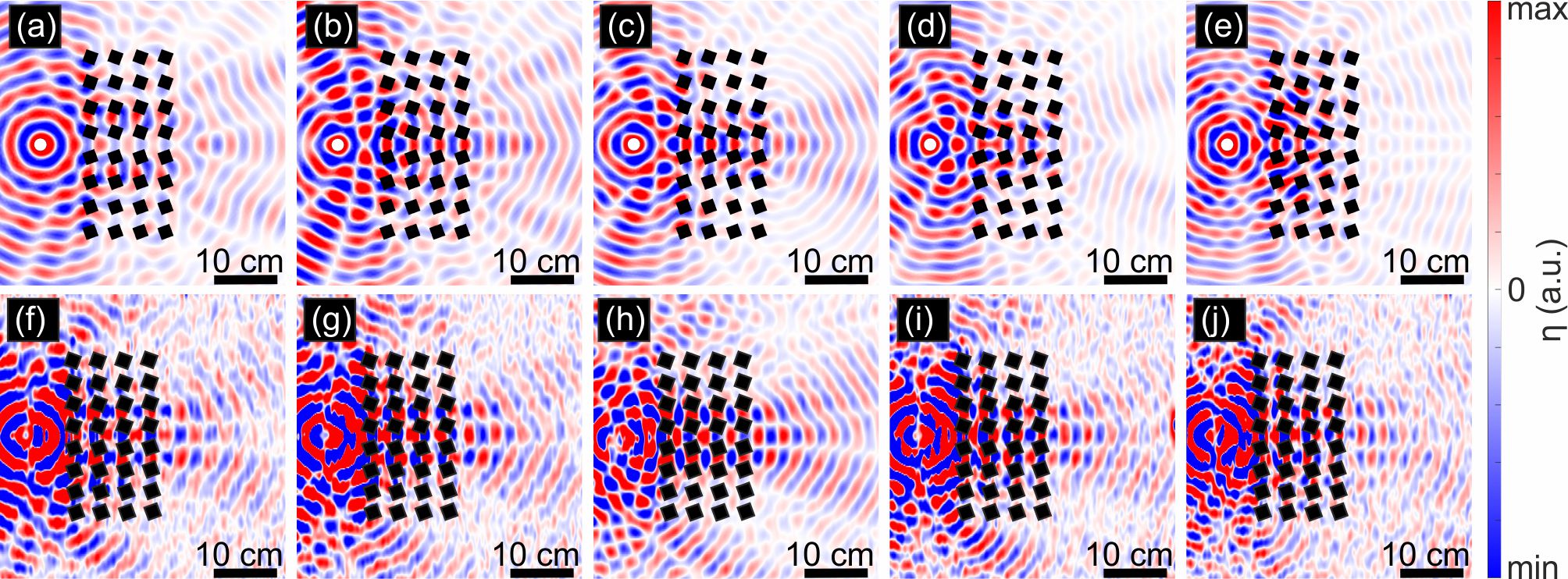}
\caption{
Valley-Hall edge states: Experiments, and simulations.
The water surface is perforated by $4\times 8$ square rigid inclusions.
Numerics (top) and experiments (bottom) are shown at frequencies (a, f) 6.7~Hz, (b, g) 7~Hz, (c, h) 7.3~Hz, (d, i) 7.6~Hz and (e, j) 7.9~Hz.}
\label{figS2}
\end{figure}

\begin{figure}[h]
\centering
\includegraphics[width=17cm]{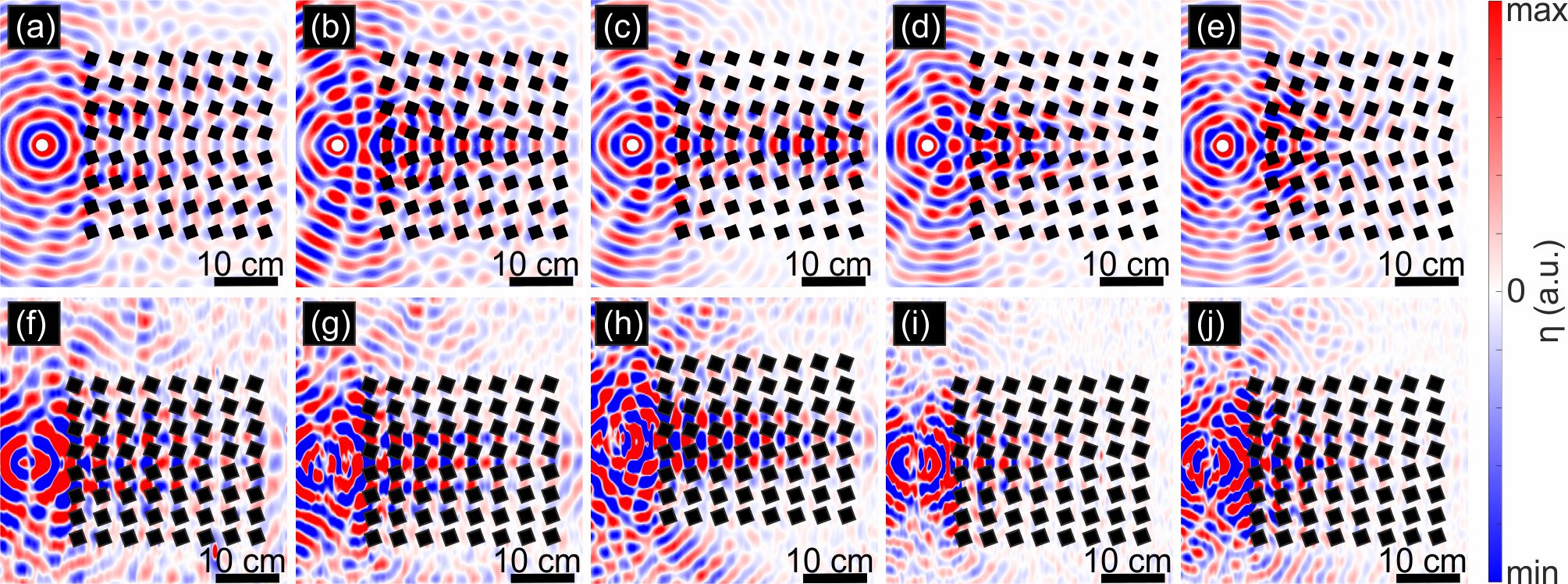}
\caption{
Valley-Hall edge states: Experiments, and simulations.
The water surface is perforated by $8\times 8$ square rigid inclusions.
Numerics (top) and experiments (bottom) are shown at frequencies (a, f) 6.7~Hz, (b, g) 7~Hz, (c, h) 7.3~Hz, (d, i) 7.6~Hz and (e, j) 7.9~Hz.}
\label{figS3}
\end{figure}

The monopolar source was used to ignite the edge states in Figs. 4, \ref{figS2} and \ref{figS3}. Figures \ref{figS2} and \ref{figS3} display further observations of topological waveguiding along a domain wall of the square lattice crystal of rotated square inclusions.
Both four (fig. \ref{figS1}) and eight crystal rows (fig. \ref{figS2}) are considered numerically and experimentally.
Frequency 6.7 Hz is below the topological band gap.
The band gap covers the 7 Hz and 7.3 Hz frequencies, at which waveguiding is observed.
Frequencies 7.6 Hz and 7.9 Hz are above the band gap.

\onecolumngrid

%

\end{document}